\documentclass[sn-mathphys-num,referee]{sn-jnl}

\usepackage{graphicx}%
\usepackage{multirow}%
\usepackage{amsmath,amssymb,amsfonts}%
\usepackage{amsthm}%
\usepackage{mathrsfs}%
\usepackage[title]{appendix}%
\usepackage{xcolor}%
\usepackage{textcomp}%
\usepackage{manyfoot}%
\usepackage{booktabs}%
\usepackage{algorithm}%
\usepackage{algorithmicx}%
\usepackage{algpseudocode}%
\usepackage{listings}%
\usepackage{caption}

\theoremstyle{thmstyleone}%
\theoremstyle{thmstyletwo}%

\theoremstyle{thmstylethree}%

\begin{document}

\title[Neural Heterogeneity Enables Adaptive Encoding of Time Sequences]{Neural Heterogeneity Enables Adaptive Encoding of Time Sequences}


\author*[1]{\fnm{Raphael} \sur{Lafond-Mercier}}\email{rlafo090@uottawa.ca}

\author[2,3,4]{\fnm{Leonard} \sur{Maler}}\email{lmaler@uottawa.ca}

\author[5]{\fnm{Avner} \sur{Wallach}}\email{avnerw@technion.ac.il}

\author[1,2,3,4]{\fnm{André} \sur{Longtin}}\email{alongtin@uottawa.ca}

\affil*[1]{\orgdiv{Department of Physics}, \orgname{University of Ottawa}, \orgaddress{\city{Ottawa}, \postcode{K1N 6N5}, \country{Canada}}}

\affil[2]{\orgdiv{Department of Cellular and Molecular Medicine}, \orgname{University of Ottawa}, \orgaddress{\city{Ottawa}, \postcode{K1H 8M5}, \country{Canada}}}

\affil[3]{\orgdiv{Brain and Mind Institute}, \orgname{University of Ottawa}, \orgaddress{\city{Ottawa}, \postcode{K1H 8M5}, \country{Canada}}}

\affil[4]{\orgdiv{Center for Neural Dynamics and Artificial Intelligence}, \orgname{University of Ottawa}, orgaddress{\city{Ottawa}, \postcode{K1H 8M5}, \country{Canada}}}

\affil[5]{\orgdiv{Faculty of Biology}, \orgname{Technion -- Israel Institute of Technology}, \orgaddress{\city{Haifa}, \postcode{3200003}, \country{Israel}}}


\abstract{Biological systems represent time from microseconds to years. An important gap in our knowledge concerns the mechanisms for encoding time intervals of hundreds of milliseconds to minutes that matter for tasks like navigation, communication, storage, recall, and prediction of stimulus patterns. A recently identified mechanism in fish thalamic neurons addresses this gap. Representation of intervals between events uses the ubiquitous property of neural fatigue, where firing adaptation sets in quickly during an event. The recovery from fatigue by the next stimulus is a monotonous function of time elapsed. Here we develop a full theory for the representation of intervals, allowing for recovery time scales and sensitivity to past stimuli to vary across cells. Our Bayesian framework combines parametrically heterogeneous stochastic dynamical modeling with interval priors to predict available timing information independent of actual decoding mechanism. A compromise is found between optimally encoding the latest time interval and previous ones, crucial for spatial navigation. Cellular heterogeneity is actually necessary to represent interval sequences, a novel computational role for experimentally observed heterogeneity. This biophysical adaptation-based timing memory shapes spatiotemporal information for efficient storage and recall in target recurrent networks.}




\maketitle

\section{Introduction}\label{intro}

Long before human-made tools~\cite{pappas_who_nodate}, biological systems devised multiple molecular and electric mechanisms to keep time over a wide range of scales~\cite{merchant_neural_2013}. Much is known about the interaction of neural activity and molecular clocks that generates hours to days long circadian rhythms~\cite{reppert_coordination_2002}. Auditory circuits in barn owls require tens of microseconds to pinpoint the direction of a sound~\cite{leibold_temporal_2001}. Scales of hundreds of milliseconds to a second, together with neural plasticity, are important for speech~\cite{poeppel_speech_2020}, song generation in birds~\cite{singh_alvarado_neural_2021}, and motor control~\cite{edwards_auditory_2002}. 

The intermediate scale of seconds to minutes is relevant e.g. to memorizing and recalling a path in space, a conversation or executing sequences of movements~\cite{buhusi_what_2005,mita_interval_2009,jazayeri_neural_2015}. An animal will memorize the time to move from a start point A to a landmark B, and from B to a target C, and thereby estimate the total time from A to C. Much effort is devoted to uncovering how the temporal structure of experience on these time scales, i.e. episodic memory, is stored for later recall and predictions of the future  ~\cite{tacikowski_human_2024,saponati_sequence_2023}. Various mechanisms across brain regions have been proposed that enable this encoding~\cite{paton_neural_2018, tsao_neural_2022}, involving clocks, activity ramping and neural sequence storage~\cite{zhou_neural_2022}. The storage of interval duration between current and future times, known as prospective timing, may engage neural timers that measure the passage of time~\cite{tsao_neural_2022}. During ramping, timing information is possibly encoded in the increasing firing rates of neurons between start and stop times. Sequence memory could use high-dimensional neural activity trajectories to represent times or time intervals. Analogs of time and ramping cells emerge in deep reinforcement learning models performing simulated interval memory tasks, but these models are agnostic to the actual mechanisms~\cite{lin_temporal_2023}.

In mammals, spatiotemporal information is available in the hippocampus, mainly area CA1 where time cells fire at given points in time, complementing place cells that fire when the animal is at given points in space~\cite{eichenbaum_time_2014}. Ongoing work seeks to explain how this information is integrated to produce sequences of time and place-specific activations of neural ensembles over hours and days~\cite{rubin_hippocampal_2015, haimerl_internal_2019}. Time cells can in principle represent intervals between contiguous events~\cite{itskov_cell_2011,macdonald_hippocampal_2011, eichenbaum_time_2014}, although the associated encoding and decoding mechanisms are not well understood. There is also evidence that representation of time duration starts before the hippocampus in sensory cortex~\cite{reinartz_direct_2024}.

A new paradigm of time coding has recently been identified in a thalamic structure in weakly electric fish known as the preglomerular complex (PG). It projects to dorso-lateral pallium~\cite{wallach_time-stamp_2018}, a likely homologue of the mammalian hippocampus~\cite{rodriguez-exposito_goldfish_2017}. PG cells use a counter-intuitive, starkly different algorithm and biophysical substrate, offering a new window into time coding in mammals as they have similar structures. PG cells receive electrosensory information about their murky nocturnal environment from the optic tectum~\cite{giassi_organization_2012-1}. The electrosense has much in common with vision~\cite{clarke_contrast_2015}: the electroreceptors in the skin read out perturbations to the fish's self-generated electric field caused by objects in the environment~\cite{jun_active_2016}. Whenever the fish encounters an object, many PG cells produce brief bursts of firings, regardless of which body part came closest to the object. The number of spikes in a burst depends on the time since the last encounter, and thus the sequence of bursts can be mapped to the sequence of intervals between encounters~\cite{wallach_time-stamp_2018}.

Adaptation has been extensively studied for various stimuli~\cite{wark_sensory_2007} and on a wide range of time scales~\cite{drew_models_2006}. For example, stimuli activating visual circuitry give rise to adaptation to contrast, orientation, or motion of the stimulus~\cite{kohn_visual_2007,clifford_visual_2007,webster_visual_2015}. Closely related paradigms attempt to explain the role of adaptation~\cite{weber_coding_2019}, such as efficient coding, by which neurons optimize the representation of stimuli given limited resources~\cite{tring_power_2023}. Another such paradigm is predictive coding, by which neural systems need to make inferences about possible future stimuli~\cite{mao_adaptation_2025}. This type of adaptive response is implicitly dependent on the time elapsed between stimuli. Here, we present an adaptation mechanism which explicitly encodes the time elapsed between identical stimuli, i.e. encounters with surrounding objects.

This elegant, tantalizing yet puzzling mechanism relies on deterministic and stochastic dynamics. Here we develop its general theory. We specifically address how it can optimally handle the diversity of prior distributions of time intervals, and the computational benefits of the strong heterogeneity of parameters across PG cells. Two-thirds of adaptive PG cells are sensitive to the history of recent encounters, and accordingly, have recovery time constants of tens of seconds. The other third can encode only the last interval. Why are both present? These questions relate to the fundamental problem of optimal information processing in the face of available biophysical diversity. Heterogeneity has attracted much attention recently as an essential part of coding schemes~\cite{gast_neural_2024,mejias_optimal_2012,marsat_cellular_2012}. Our work presents a deep analysis of these questions, and proves that heterogeneity is in fact required in order to properly encode more than one interval in a sequence. 

We first explain the adaptive stochastic firing model and the Bayesian estimator of time intervals from PG burst sizes. We then quantify its performance using Fisher information. This quantification shows that, for the single interval case, the optimal decoder should have memory-less cells (a third of the cells in fish) that respond only to the last time interval. The effect of the diversity in time constants is then studied for encoding a single interval. We then generalize our results by considering a continuous prior distribution of intervals and of PG time constants. Although a homogeneous network appears optimal for single interval estimation, a suitably chosen heterogeneous network still performs quite well. This single interval property is advantageous as we end by showing that heterogeneity is actually necessary for estimating multiple intervals. Hence, heterogeneity can be leveraged for a complex task without hindering the latest interval estimates significantly.

\section{Results}\label{results}

We consider a network of independent cells receiving brief input stimulation in parallel at times $t_k = \sum\limits_{i=1}^{k} T_i$, where $T_i$ is the time interval between encounters happening at times $t_{i-1}$ and $t_i$, with $t_0 = 0$ the time of the first stimulus. We model the dynamical adaptation of PG cells as in~\cite{wallach_time-stamp_2018} using a simple resource based model~\cite{tsodyks_neural_1997} with long replenishment time constants. A cell has an amount of resources available to produce spikes during an encounter with an object which is represented by the variable $x$. During an encounter, the resource variable is depleted by an amount that depends on the memory parameter $\beta$, and then recovers at a rate $\tau$ (see Figure~\ref{fig:dynamical model}A and Equation~\eqref{eq:resource variable}). The value of this resource variable during an encounter is then passed through a linear rectification with gain $a$ and baseline activity $c$ to generate a firing parameter $\lambda$ (see Figure~\ref{fig:dynamical model}B and Equation~\eqref{eq:firing parameter}). Finally, this firing parameter is fed to a Poisson probability distribution that represents the probability of having a given spiking response from this particular cell during the encounter (see Figure~\ref{fig:dynamical model}C and Equation~\eqref{eq:spike distribution}). We then use samples from this Poisson distribution to simulate the number of spikes that would be observed in PG cells during an encounter. The process is summarized in the following equations:
\begin{align}
    x_n\left( \{T_i\}_{i=1}^{n} \right) &= 1 - e^{-T_n/\tau}(1 - \beta x_{n-1}),\label{eq:resource variable}\\
    \lambda_n\left( \{T_i\}_{i=1}^{n} \right) &= \left[ ax_n + c \right]_+,\label{eq:firing parameter}\\
    P\left(R_n\right) &= \frac{(\lambda_n)^{R_n}e^{-\lambda_n}}{R_n!}\label{eq:spike distribution}.
\end{align}

From now on, the subscript of a variable indicates the index of an encounter while the superscript indicates the index of an adaptive neuron. Parentheses will be used to differentiate indices from exponents. These neurons project downstream to a decoder that extracts the sequence $\{T_i\}$ for further use (such as storage, not modeled here). We approach this problem using a Bayesian framework combined with signal detection theory and simple neuron stochastic dynamics with or without memory. Neuron output spikes thus result from a stochastic spike-generating process, with time-dependent parameter set by single-neuron Markovian deterministic dynamics that are driven by external inputs, namely, the sequence of intervals between encounters. Recovering the sequence of time intervals from the responses of PG cells can be achieved through maximum likelihood estimation (MLE), which is an efficient estimator given enough cells. From Bayes' rule, we write the probability of some sequence of time intervals $\{T_i\}$ to have generated a set of $N$ responses $\{R_n^j\}_{j=1}^{N}$ during the $n$-th encounter as

\begin{equation}
    P\left( \{T_i\}_{i=1}^n | \{R_n^j\}_{j=1}^{N} \right) = \frac{P\left( \{R_n^j\} | \{T_i\} \right)P\left( \{T_i\} \right)}{P\left( \{R_n^j\} \right)}.
    \label{eq:bayes rule}
\end{equation}

Finding the set of intervals $\{T_i\}$ that maximizes Equation~\eqref{eq:bayes rule} yields the maximum a posteriori estimator. This estimator makes use of prior information on the sequence of time intervals through $P\left(\{T_i\}\right)$. However, since we want to quantify how much of the sequence of time intervals is encoded solely in the response of PG neurons, we need to assume no prior information is available for the estimation. Maximizing Equation~\eqref{eq:bayes rule} then becomes equivalent to maximizing the likelihood

\begin{equation}
    L\left( \{T_i\}_{i=1}^n | \{R_n^j\}_{j=1}^{N} \right) = P\left( \{R_n^j\}_{j=1}^{N} | \{T_i\}_{i=1}^{n} \right).
    \label{eq:likelihood}
\end{equation}

It is numerically easier to maximize the logarithm of the likelihood. Therefore, for an ensemble of $N$ independent adaptive neurons, we maximize

\begin{align}
    \ell \left( \{T_i\}_{i=1}^{n} | \{R_n^j\}_{j=1}^{N} \right) = \sum_{j=1}^{N}& \left( R_n^j \ln \lambda_n^j\left(\{T_i\}_{i=1}^{n}\right) -\right. \nonumber \\ & \left. \lambda_n^j\left(\{T_i\}_{i=1}^{n}\right) - \ln R_n^j ! \right)
    \label{eq:loglikelihood}
\end{align}

\noindent to retrieve the most likely sequence $\{T_i\}$ resulting from the observed response (see Figure~\ref{fig:dynamical model}D).

\begin{center}
\begin{figure}[H]
    \includegraphics[width=\linewidth]{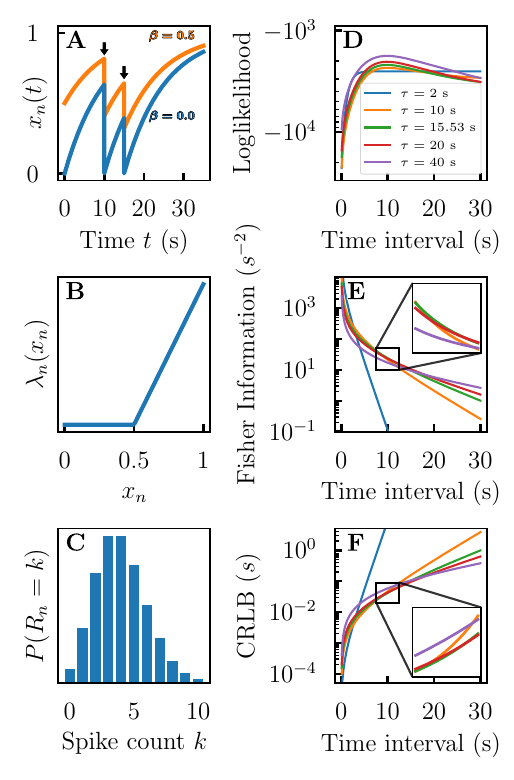}
\end{figure}
\captionof{figure}{\textbf{Generating the adaptation responses, estimating time interval and quantifying the performance of maximum likelihood estimation.} (A) A resource variable $x$ recovers with time constant $\tau$ and is depleted by some amount during an encounter (black arrows) depending on the memory parameter $\beta$ (see Equation~\eqref{eq:resource variable}). (B) During each encounter, a firing parameter $\lambda$ is computed from the resource variable $x$ through a rectified linear transformation with gain parameter $a$ and baseline activity $c$ (see Equation~\eqref{eq:firing parameter}). (C) A Poisson distribution with parameter $\lambda$ dictates the number of spikes generated during an encounter (see Equation~\eqref{eq:spike distribution}). (D) Example of the loglikelihood for a real time interval of $10$~s between encounters, computed with the responses of $1,000$ identical model PG cells during the last encounter. This is shown for five recovery time constants $\tau$. The log-likelihood has a maximum around $10$~s for most populations (see Equation~\eqref{eq:loglikelihood}). (E) Fisher information (FI) contained in the response of $1,000$ PG cells for the same set of $\tau$ values (see Equation~\eqref{eq:fisherinfo}). (F) Cramér-Rao Lower Bound (CRLB) of the maximum likelihood estimator using the response of $1,000$ identical PG cells for the same set of $\tau$ values (see Equation~\eqref{eq:CRLB}). We see that the FI (CRLB) decreases (increases) as $T$ increases, and that a value of $\tau$ gives larger (smaller) FI (CRLB) than another one only in a specific range of values of $T$. (D) to (F) $a = 10$, $c = 0$ and $\beta = 0$.}
\label{fig:dynamical model}
\end{center}

\subsection*{Cell parameters can be optimized through Fisher information}

We first assess how the parameters of the network affect the time interval estimation. To do so, we use the Fisher information (FI), which is a measure of maximum precision of an unbiased estimator. Indeed, although not exactly the same as an information theoretic type of information, it is used to compute a lower bound of the mean squared error (MSE) made by the estimator called the Cramér-Rao lower bound (CRLB, see Equation~\eqref{eq:CRLB}).

Since the responses of PG neurons during an encounter are assumed to be independent Poisson variables, we can add the FI of the individual neurons to get the total FI of a network of PG cells (see Equation~\eqref{eq:fisherinfo})~\cite{clarke_neural_2015}. The resulting expression depends on the values of the time intervals in the stimulus sequence (see Figure~\ref{fig:dynamical model}E-F) as well as on the different gain ($a^k$), baseline activity ($c^k$), memory ($\beta^k$) and recovery time ($\tau^k$):

\begin{align}
    \left[\mathcal{I}\right]_{ij} &= \sum_{k=1}^{N} \frac{1}{\lambda_n^k}\left( \frac{\partial\lambda_n^k}{\partial T_i} \right)\left( \frac{\partial\lambda_n^k}{\partial T_j} \right) \label{eq:fisherinfo},\\
    \left[CRLB\right]_{ij} &= \left[ \mathcal{I}^{-1} \right]_{ij} \leq \left[MSE\right]_{ij} \label{eq:CRLB}.
\end{align}

To achieve a comprehensive understanding of this dynamical system, we begin by focusing on the simplest case of a single time interval $T$ between two consecutive encounters. We neglect previous encounters by setting the initial state of the neurons $x_0 = 1$, which they eventually reach when there has not been an encounter for a significant time. Optimizing the FI for this specific case is simple and can be done by looking at the partial derivatives of the FI with respect to the different cell parameters (see Methods). We show that, for any value of the time interval $T$, the partial derivative with respect to the gain parameter $a$ is always positive, which means that a gain parameter as large as possible is preferable (see Figure~\ref{fig:optimization of estimator}A). Similarly, the derivative with respect to the memory parameter $\beta$ and the baseline activity $c$ are always negative; an indication that having no memory ($\beta = 0$) and no spontaneous activity ($c = 0$) optimally encodes a single time interval (see Figure~\ref{fig:optimization of estimator}B-C).

The situation is different for the derivative with respect to $\tau$, where a single value of $\tau$ maximizes the FI for a specific choice of $T$ (see Figure~\ref{fig:optimization of estimator}D). We numerically compute through Newton's root finding algorithm (see Methods) that $\tau \approx 1.5533 T$ optimizes the FI when $\beta = c = 0$. However, the network should not be optimized to estimate a singular time interval value. Rather, it should be able to optimally estimate the wide range of interval values found in nature. To do so, we can look at the expectation value of the CRLB for a given prior of time intervals. For a single population of $N$ cells, gain $a$, time constant $\tau$ and without baseline activity or memory ($c = \beta = 0$), we can explicitly write

\begin{equation}
    E_T\left[\text{CRLB}\right] = \frac{\tau^2}{aN}\left\{M_T\left(\frac{2}{\tau}\right) - M_T\left(\frac{1}{\tau}\right)\right\},
    \label{eq:moment generating expectation}
\end{equation}

\noindent where $E_T[\cdot]$ is the expectation value with respect to $T$ and $M_T(x) = E_T\left[e^{xT}\right]$ is the moment-generating function dependent on the chosen prior distribution of time intervals. It is then possible to find the value of $\tau$ that minimizes this average CRLB. For example, an exponential prior distribution of time intervals with an average of $\hat{T}$ yields an optimal time constant of $\tau = (3 + \sqrt{3})\hat{T} \approx 4.73\hat{T}$, while a uniform distribution between $0$ and $T_\text{max}$ yields an optimal time constant of $\tau \approx 1.14 T_\text{max}$. Moreover, we show that for any prior distribution of time intervals, there can be at most one value of $\tau$ which minimizes the average CRLB (see Supplementary Information). However, it is not clear what the effect of having more populations of cells with different time constants is on the error. This therefore raises the question of what distribution of time constants is optimal given a time interval prior distribution.

\begin{center}
\begin{figure}[H]
    \includegraphics[width=\linewidth]{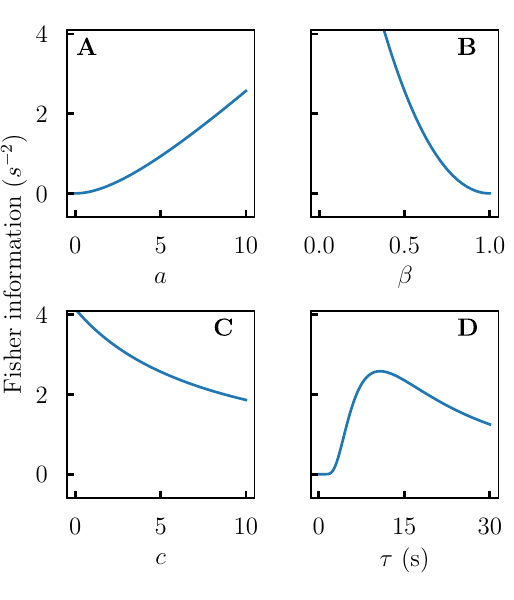}
\end{figure}
\captionof{figure}{\textbf{Maximizing the Fisher information of the maximum likelihood estimator of a single time interval for the homogeneous case.} Example of the FI (see Equation~\eqref{eq:fisherinfo}) for $N=1,000$ cells and an interval of $10$~s as a function of the (A) gain $a$, (B) memory $\beta$, (C) baseline activity $c$ and (D) recovery time $\tau$. Maximum values of FI are reached at either end of the parameter domain with the exception of $\tau$. Unless otherwise stated, the values of the parameters are $a=10$, $c=5$, $\tau = 10$~s, $\beta = 0.5$ and $x_0 = 1$. Note that the maximum of $\tau$ in (D) would be at $15.5$~s if $\beta$ and $c$ were zero.}
\label{fig:optimization of estimator}
\end{center}

\subsection*{Heterogeneity of time constants has a small negative impact on single interval estimates}

To determine whether combining different values of $\tau$ over different cells is beneficial in the estimation of the time interval values of interest, we start by building our intuition using six simplified cases. For all six cases, we choose fixed parameters for $a$, $c$ and $\beta$ while looking at the effect of $\tau$. The first two cases looks at the CRLB of a network of $1,000$ cells with a unique value of $\tau^1$ and $a^1 = 5$. When looking at the CRLB at a specific value of time interval $T=10$~s (see Figure~\ref{fig:optimization of recovery parameter}A), we see a unique maximum at $\tau^1 \approx 15.5T$, which is expected because of the relationship between $\tau$ and $T$ that was previously mentioned. We then average the CRLB over two distinct values of time intervals $T = 10$~s and $T = 15$~s (see Equation~\eqref{eq:CRLB comparison}). We find that no new maximum appears and that the optimal value of $\tau^1$ is simply shifted (see Figure~\ref{fig:optimization of recovery parameter}B). 

Next, the same computations are repeated after adding a second, possibly different value of $\tau$. In other words, we have a network composed of a sub-population of $500$ cells with $\tau^1$ and $a^1 = 5$, and another sub-population of $500$ cells with $\tau^2$ and  $a^2 = 15$. In the case where we compute the CRLB for a single time interval $T = 10$~s, the optimal combination of $\tau^1$ and $\tau^2$ is unique, namely, $\tau^1 = \tau^2 \approx 15.5T$ (see Figure~\ref{fig:optimization of recovery parameter}D). This is also the case when averaging the CRLB over two distinct time intervals, where the optimal value of $\tau^1 = \tau^2$ is shifted, (see Figure~\ref{fig:optimization of recovery parameter}E). To make sure this is not simply due to the similarity between both intervals, we repeat the same computation with $T = 2$~s and $T = 20$~s (see Figure~\ref{fig:optimization of recovery parameter}C,F). The observations are the same even with this large difference in time interval values. These simple cases suggest that a homogeneous distribution of $\tau$ may actually be optimal if we want to maximize the average performance of the MLE for a single time interval. 

Repeating the same exercise for the relative CRLB (dividing the error by the time interval value), the results are similar with the exception of new maxima being added when shorter time intervals are considered (see Figure S1F). The old maximum (see Figure S1C) becomes a saddle point when a new value of $\tau$ is added.

\begin{center}
\begin{figure}[H]
    \includegraphics[width=\linewidth]{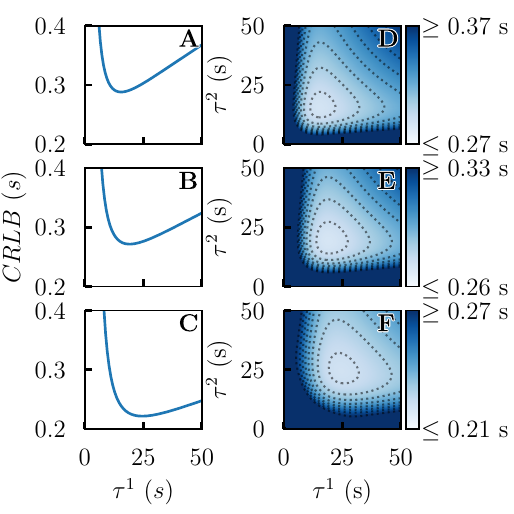}
\end{figure}
\captionof{figure}{\textbf{Six specific cases to assess how adding new sub-populations affects the error made on single time interval estimates.} (A) to (C) Single population with only one recovery parameter $\tau^1$, gain $a^1=5$, baseline activity $c^1=0$ and memory parameter $\beta^1 = 0$. (A) Cramér-Rao lower bound (CRLB) for a time interval of $T = 10$~s. (B) Average of the CRLB (see Equation~\eqref{eq:CRLB comparison}) for time intervals of $T = 10$~s and $T = 15$~s. The minimum shifts slightly towards a larger value of $\tau^1$ when compared to (A). (C) Average of the CRLB for time intervals of $T = 2$~s and $T = 20$~s. The minimum remains unique even with a large difference between time interval values. (D) to (F) Two populations with recovery parameters $\tau^1$ and $\tau^2$, gain parameters $a^1 = 5$ and $a^2 = 15$, baseline activity $c^1 = c^2 = 0$ and memory parameter $\beta^1 = \beta^2 = 0$. (D) CRLB for a time interval of $T = 10$~s. (E) Average of the CRLB for time intervals of $T = 10$~s and $T = 15$~s. Similarly to (B), the minimum shifts slightly towards larger values of $\tau^1$ and $\tau^2$. (F) Average of the CRLB for time intervals of $T = 2$~s and $T = 20$~s. The minimum remains unique even with a large difference between time interval values. For (D) to (F): dashed lines represent contours with the same value of CRLB and the minimum is on the diagonal $\tau^1 = \tau^2$ with the same value as in (A) to (C), respectively.}
\label{fig:optimization of recovery parameter}
\end{center}

\begin{figure}[H]
    \centering
    \includegraphics[width=\linewidth]{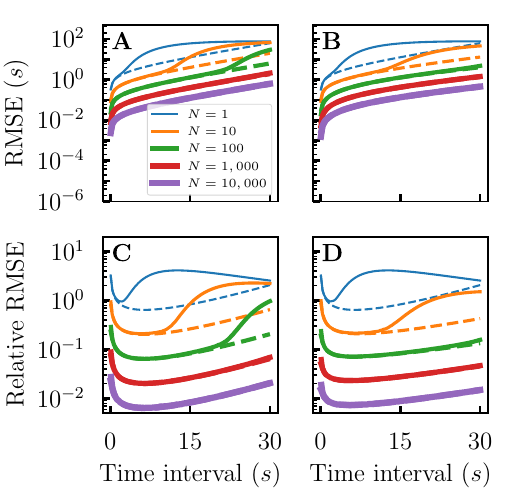}
    \caption{\textbf{The error tends towards the Cramér-Rao lower bound for a single interval.} Monte Carlo computation of the root mean square error (RMSE, solid lines, see Equation~\eqref{eq:RMSE}) and Cramér-Rao lower bound (CRLB, dashed lines, see Equation~\eqref{eq:general discrete CRLB}) of networks with increasing cell counts as a function of the value of the time interval between encounters. (A) Homogeneous network with recovery time parameter $\tau = 10$~s for all cells. There is a significant bias causing the RMSE to diverge from the CRLB when the cell count is low. (B) Heterogeneous network with $\tau$~sampled from a uniform distribution between $0.1$~s and $20$~s. There is also a bias for this heterogeneous network, but it is reduced more rapidly when increasing the number of cells than in the homogeneous case. (C) and (D) Same as (A) and (B) respectively, but the relative error $\text{RMSE}/T$ is shown. For both networks, $a = 10$, $c = 0$ and $\beta = 0$.}
    \label{fig:RMSE vs CRLB}
\end{figure}

To assess the validity of this result, we compare the actual error made with MLE with the CRLB. In the case of a sequence of only one time interval, the root mean square error (RMSE) computed through Monte Carlo simulations tends towards the CRLB (see Figure~\ref{fig:RMSE vs CRLB}A,B). For a network comprised of only $1,000$ cells, the bias is practically non-existent, and the RMSE is extremely close to the CRLB. This justifies the use of FI for the parameter optimization of the MLE. There is a difference in the behavior of the RMSE between a homogeneous network (see Figure~\ref{fig:RMSE vs CRLB}A) and a random heterogeneous network (see Figure~\ref{fig:RMSE vs CRLB}B). That is, the error made due to the bias decreases much faster as $N$ increases in the heterogeneous case than in the homogeneous one. \textit{This suggests that heterogeneity helps in a situation where bias is important.} However, since PG contains approximately $60,000$ cells~\cite{trinh_cryptic_2016}, we decided not to explore this effect further.

To see if the trends observed with the six previously mentioned specific cases hold in general, we compute the CRLB over a wide range of $\tau$ and $T$ values (see Figure~\ref{fig:RMSEaverages}A). At first glance, we see that increasing $T$ also increases the error made on the estimates monotonically (see Figure~\ref{fig:RMSEaverages}B). For a specific $T$, increasing $\tau$ initially has the effect of decreasing the error significantly until a minimum is reached. The increase after this minimum is then extremely slow, suggesting similar behavior is expected from a large non optimal $\tau$ (see Figure~\ref{fig:RMSEaverages}C).

We then look at what happens when introducing a continuous distribution of values for the recovery times in a network, $P_\tau(\tau)$, combined with a continuous weight (or "prior") function for the values of the time intervals of interest $P_T(T)$. Specifically, we make the weight function a power law (see Equation~\eqref{eq:powerlaw}) and take the distribution of recovery times to be log-normal~\cite{gussin_limits_2007, buzsaki_log-dynamic_2014} 
(see Equation~\eqref{eq:lognormal}). The power law is described by the exponent $k$ where a negative value gives more weight to larger time intervals while a positive value gives more weight to smaller time intervals. A value of $k=0$ gives equal weight to all time interval values in the domain of interest $[T_\text{min} ; T_\text{max}]$ (see Figure~\ref{fig:continuous distribution}A). As for the distribution of $\tau$, it is described by mean $\mu$ and variance $\sigma^2$ (see Figure~\ref{fig:continuous distribution}B). The distributions are defined by

\begin{align}
    P_T(T) &= \alpha T^{-k} \label{eq:powerlaw},\\
    P_\tau(\tau) &= \frac{1}{\tau s \sqrt{2\pi}}\exp\left(-\frac{(\ln\tau - m)^2}{2s^2}\right) \label{eq:lognormal},
\end{align}

\noindent where $\alpha$ is such that $\int_{T_\text{min}}^{T_\text{max}} P(T)\text{d}T = 1$, $m = \ln(\mu^2/\sqrt{\sigma^2 + \mu^2})$ and $s = \ln(1 + \sigma^2/\mu^2)$ with $\mu$ and $\sigma^2$ being the mean and variance of the $\tau$ distribution, respectively.

The variance in the parameter distribution is how we introduce heterogeneity in this continuous case. To assess how it affects the performance of MLE, we first compute the CRLB using the FI of a full network containing recovery times distributed as $P_\tau(\tau)$. This allows us to get an idea of the performance of the whole network given some stimulus. The average $\mu$ of this $\tau$ distribution has the most important effect (see Figure~\ref{fig:RMSEaverages}E) on the error of the estimate, while the effect of the variance $\sigma^2$ is less noticeable (see Figure~\ref{fig:RMSEaverages}F).

\begin{center}
\begin{figure}[H]
    \includegraphics[width=\linewidth]{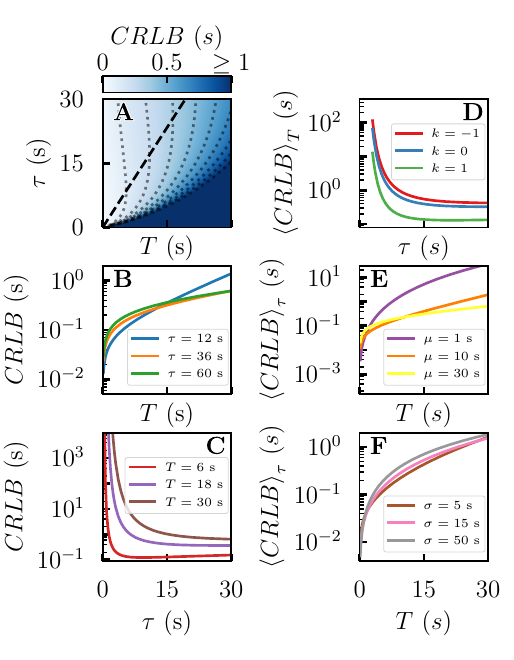}
\end{figure}
\captionof{figure}{\textbf{Cramér-Rao lower bound and its averages for a single interval estimated by a homogeneous VS heterogeneous population.} \textbf{a-d}: Homogeneous network with a single recovery parameter $\tau$. (A) Dotted lines: contour lines with identical CRLB value. Dashed line: relationship between a specific value of $T$ and the associated optimal value of $\tau$ (see Equation~\eqref{eq:dFIdtau}). (B) Cross sections of the CRLB for various values of $\tau$ as a function of $T$. The error is lower for smaller values of $\tau$ when looking at small time intervals, but it quickly grows thereafter (eg. blue vs. orange curves). (C) Cross sections of the CRLB for various values of $T$ as a function of $\tau$. The overall error is lower for smaller intervals, and the associated optimal value of $\tau$ increases with the length of the interval. (D) Averaged CRLB of a homogeneous network where the values of the time intervals are power-law distributed (see Equation~\eqref{eq:powerlaw}). (E) CRLB of heterogeneous networks whose recovery time parameters are log-normal distributed (see Equation~\eqref{eq:lognormal}) with constant $\sigma=1$~s and different values of $\mu$ as a function of $T$. (F) Same as (E), but with constant $\mu = 10$~s and different values of $\sigma$.}
\label{fig:RMSEaverages}
\end{center}

Since there is a wide range of different possible stimuli, we settle on the average of the CRLB sampled from the $P_T(T)$ distribution as a metric to measure the overall performance of the network over this range of time intervals of interest. The effect of averaging over different time intervals seems to be a flattening of the CRLB around the minimum, making large values of $\tau$ perform essentially equally, especially when giving more weight to large time intervals (see Figure~\ref{fig:RMSEaverages}D). We then combine both averagings to get the overall effect of continuous distributions which yields the expression for the average CRLB:

\begin{equation}
    \overline{\text{CRLB}} = \int_{T_\text{min}}^{T_\text{max}} \frac{P_T(T)}{\sqrt{\int_{0}^{\infty} P_\tau(\tau) \mathcal{I}(T; \tau) \text{d}\tau}}\text{d}T.
    \label{eq:avgCRLB}
\end{equation}

We can also compute the average relative CRLB in a similar manner:

\begin{equation}
    \text{rel-}\overline{\text{CRLB}} = \int_{T_\text{min}}^{T_\text{max}} \frac{P_T(T)}{T\sqrt{\int_{0}^{\infty} P_\tau(\tau) \mathcal{I}(T; \tau) \text{d}\tau}}\text{d}T.
    \label{eq:avgrelCRLB}
\end{equation}

We then find the averages $\mu$ that minimize these quantities for different values of $\sigma$ and $k$ (see Figure~\ref{fig:continuous distribution}C,E). The values of the minima reached during the optimization of Equations~\eqref{eq:avgCRLB} and \eqref{eq:avgrelCRLB} are then used to compare the minima for all specified values of $\sigma$ and $k$ (see Figure~\ref{fig:continuous distribution}D,F). The smallest possible average CRLB is reached when there is a single value of $\tau$ (lower black dashed line) for all values of $k$ between $-1$ and $1$. However, when optimizing the relative CRLB, heterogeneity is better when shorter time intervals are of greater importance ($k \sim 1$). The improvement is small, but noticeable (see Figure \ref{fig:continuous distribution}F). This could be due to specific combinations of time interval values in the prior distribution. More specifically, it seems more important when shorter time intervals are included (see Figure S1F). Homogeneity is therefore the optimal solution for estimating a sequence of a single time interval in most cases. This goes in accordance with the previously built intuition using six specific cases. 

However, it should be noted that the actual difference in CRLB is not that significant between a network that is homogeneous in $\tau$ vs one that has a well-chosen heterogeneity in $\tau$. Indeed, even for a significantly heterogeneous network with $\sigma = 16$~s, the maximum difference with the error made in the homogeneous case is only about $3\%$. It is also important to point out that not every heterogeneity is a sensible choice. For example, a completely uniform distribution of $\tau$ (black dotted line) yields quite a significant error. This is due to the large proportion of neurons with smaller time constants that recover quickly and hence contribute little to the estimation of larger time intervals. This small dependence of the error on the variance therefore means that a reasonable choice in the heterogeneity of time constants can be made almost independently from that for the time intervals expected to be experienced in nature.

\begin{center}
\begin{figure}[H]
    \includegraphics[width=\linewidth]{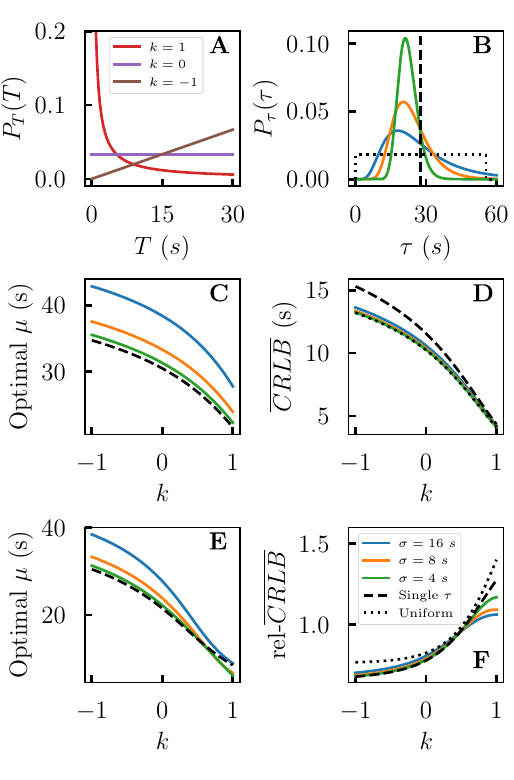}
\end{figure}
\captionof{figure}{\textbf{Finding the optimal distributions of time recovery parameters given a stimulus prior and fixed heterogeneity by minimizing the average CRLB.} (A) Power-law distributions for the continuous limit of stimulus prior (see Equation \eqref{eq:powerlaw}). This is meant to represent how time intervals are expected to be distributed in nature. (B) Optimal log-normal distributions of the recovery time parameters for $k=1$ (see Equation \eqref{eq:lognormal}). This represents the heterogeneity found in the model PG network. (C) Average of log-normal time constant distribution $\mu$ for fixed standard deviations $\sigma$ that minimizes the average CRLB given a power law prior. (D) Average of the CRLB (see Equation \eqref{eq:avgCRLB}) of the optimal distributions of $\tau$ with fixed standard deviations $\sigma$. For optimal log-normal distributions of $\tau$, the difference in performance is minimal even for a large spread of recovery times. However, a uniform distribution of $\tau$ makes the average error of the estimates significantly larger, indicating a sensible choice of $\tau$ distribution is necessary. (E) Same as (C), but $\mu$ minimizes the average relative CRLB (see Equation \eqref{eq:avgrelCRLB}). (F) Same as (D), but the average relative CRLB is minimized. Heterogeneity seems to help in the case of large $k$, i.e. when smaller values of $T$ are considered more important. For (B) to (F) Dashed line is a single time constant $\tau$ minimizing the error (homogeneous limit). Dotted line is a network with uniformly distributed $\tau$ between $0.1$ s and $56$ s ($\sigma = 16$ s). Blue: $\sigma = 16$ s, Orange: $\sigma = 8$ s, Green: $\sigma = 4$ s.}
\label{fig:continuous distribution}
\end{center}

\subsection*{Heterogeneity is necessary for estimating sequences of multiple time intervals}

To provide an explanation for the observed heterogeneity in real PG neurons of electric fish, we need to look at the case where a sequence of encounters produces 2 or more time intervals. The simplest such case is when there are 2 time intervals to be estimated with 2 different populations of neurons. In this case, the FI becomes a $2 \times 2$ matrix whose inverse gives the CRLB. For this matrix to be invertible, its determinant must be non-zero. For these 2 types of neurons, the expression of the determinant of the FI matrix can be reduced quite significantly (see Methods): 

\begin{equation}
    \text{det}\mathcal{I} = \frac{N^1a^1N^2a^2}{\lambda_2^1\lambda_2^2}\left(\frac{\partial x_2^1}{\partial T_1}\frac{\partial x_2^2}{\partial T_2} - \frac{\partial x_2^1}{\partial T_2}\frac{\partial x_2^2}{\partial T_1}\right)^2.
    \label{eq:FIdet}
\end{equation}
The only way for the determinant to be non-zero is for both types of cells to have either a different value of recovery time $(\tau^1 \neq \tau^2)$ or a different value of memory parameter $(\beta^1 \neq \beta^2)$. Another requirement is to have at least one type of cell that can encode the first time interval in the sequence, i.e. one population with $\beta > 0$. A different gain $(a^1 \neq a^2)$ or a different baseline activity $(c^1 \neq c^2)$ cannot make this determinant non-zero by themselves. Thus, a surprising result arises from our analyses: at least some level of heterogeneity is needed to encode two or more intervals, as opposed to the case of a single time interval.

Although a non-zero determinant is a necessary condition to effectively retrieve two time intervals from the responses of two populations of neurons, it does not guarantee a unique solution every time. Indeed, there can be a combination of time intervals and responses for which the log-likelihood (LL) has a non-unique maximum. For example, if the average response of a population is larger than $a + c$, the estimate diverges ($T \to \infty$). This makes it impossible to retrieve exactly what the time intervals actually were. However, due to the intrinsic variability of the response of adaptive neurons, the exact same time intervals could later yield a LL with a unique maximum. This happens when two populations with memory are similar. The determinant of the FI matrix is non-zero, but the MLE might not work every time depending on the response of adaptive cells. This rarely happens for large populations of neurons, because the averaging effect over multiple cells restrains the possible value of the average response of a population. 

\begin{figure}[H]
    \centering
    \includegraphics[width=\linewidth]{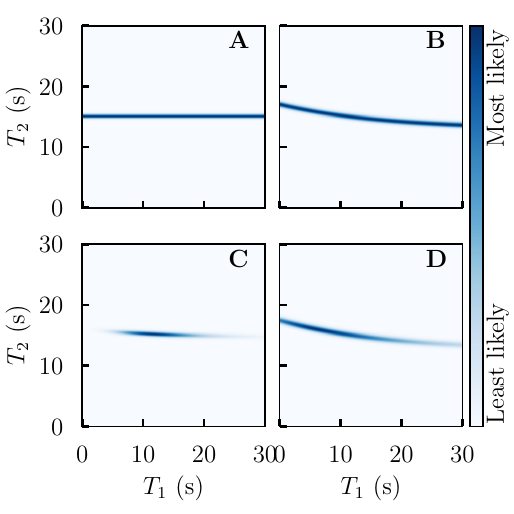}
    \caption{\textbf{Inferring two time intervals with one or two populations of neurons.} (A) to (D) Likelihood of the values of a sequence of two time intervals from the response of adaptive cells during the last encounter. The stimuli were a time interval of $10$~s ($T_1$) followed by a time interval of $15$~s ($T_2$). (A) Single population of $1,000$ adaptive neurons with no memory ($\beta = 0$). (B) Single population of $1,000$ adaptive neurons with memory ($\beta = 0.3$). (C) Two populations of $500$ adaptive neurons each. One population has memory ($\beta = 0.3$) while the other does not ($\beta = 0$). (D) Two populations of $500$ adaptive neurons each. Both populations have memory ($\beta_1 = 0.3$, $\beta_2 = 0.5$). All points on the lines in panels (A) and (B) are equally likely, which indicates at least 2 different populations are necessary to have a single maximum. The maximum in (C) is more pronounced than the one in (D), which indicates that having a population without memory yields more precise estimates.}
    \label{fig:likelihood example}
\end{figure}

Moreover, when combining a population with memory with one without, this phenomenon is impossible. This is because the memoryless population gives a unique solution for the latest time interval, which in turn gives only one value for the previous interval that maximizes the LL. In fact, the combination of two populations where one of them is memoryless maximizes the determinant of the FI, possibly minimizing the error made on the estimates of the total time travelled $T_1 + T_2$. We can illustrate this phenomenon by looking at the LL of different combinations of populations (see Figure \ref{fig:likelihood example}). The maximum of the LL looks sharpest when one of the populations has no memory. This may therefore explain why a large fraction of the cells were observed to be memoryless.

\section{Discussion}\label{sec12}

\subsection*{Dynamical Bayesian Framework for Adaptive Encoding}

We presented a joint dynamical and Bayesian analysis of interval encoding, focusing first on a single time interval between two encounters, and ending with a proof that heterogeneity is needed to encode two or more intervals. MLE combined with Fisher and Cramér-Rao metrics allowed us to quantify how well intervals are represented with the adaptive time-stamp mechanism given the strong parametric variability, especially the presence of cells with and without memory. The choice of MLE was mostly made for the sake of simplicity. It assumes minimal knowledge about how PG cells behave and what kind of prior information electric fish might have about the distribution of time intervals encountered in their natural habitat. Here we have extended the MLE to include such priors, parametric heterogeneity and memory effects.

We do not claim that downstream structures from PG actually implement MLE, though some sort of Bayesian interpretation could be applied~\cite{ma_bayesian_2006,deneve_bayesian_2008}. In fact, the downstream dorsolateral region (DL) and neighboring areas are highly recurrent~\cite{giassi_organization_2012}. It is therefore suspected that interval sequence information can actually be stored in an attractor network, and that more accurate information about sequences may be available than our theory suggests. The FI metric represents how much information an efficient estimator has about the sequence of time intervals using only the response during the latest encounter and is therefore independent of the actual method used to decode such information.

Different schemes have been introduced to explain the encoding of time in neural networks, such as state-dependent networks~\cite{karmarkar_timing_2007}, pulse-counting~\cite{zemlianova_biophysical_2022}, oscillator-based models~\cite{matell_cortico-striatal_2004}, and sequences of neuronal assemblies~\cite{itskov_cell_2011,rubin_hippocampal_2015,haimerl_internal_2019}. Our time-stamping mechanism resembles ramping activity models~\cite{simen_model_2011} in that neuronal firing probability builds over time since an event. However, the biophysical origin of its long recovery time scale is unknown. Nevertheless, the relationship between time interval estimation and spatial learning is evident. A multitude of strategies have been put forward for spatial learning in teleost fish alone~\cite{rodriguez_spatial_2021}, for which sensory information needs to be combined and then stored in DL, much like what happens in the mammalian hippocampus. Path integration, the use of sensory information to estimate distance travelled, likely benefits from downstream combinations of information about intervals, place, heading direction, as well as velocity from the lateral line organs~\cite{wallach_time-stamp_2018}, making the activity of this thalamus-like structure crucial for spatial learning.

\subsection*{Decoding a single interval}

For a single interval, it is best to have no memory beyond the most recent interval ($\beta = 0$), nor baseline activity ($c = 0$), and for the gain $a$ to be as large as possible. Further, for any specific value of the current time interval $T$ to be estimated from neural activity, there is an optimal recovery time constant $\tau \approx 1.55T$. A similar kind of result can be found for any distribution of time interval values. This prompted the exploration of heterogeneity in $\tau$. We hypothesized that the optimal parameters should be those that minimize the average CRLB over the interval prior. The Monte Carlo simulation of the estimation error revealed how a large number of neurons enables CRLB-based optimization. 

We initially introduced heterogeneity using six specific cases of combinations of $T$ and $\tau$ to build intuition about the role of heterogeneity. These cases suggested that a homogeneous network was best for encoding a single time interval. This was further confirmed in the more general setting where network heterogeneity is modeled by a log-normal distribution, and the CRLB is averaged over a power-law interval prior. A suitably chosen heterogeneous network can also have a relatively small average CRLB for a single interval, a desirable property that clearly needn't be sacrificed for multiple interval estimation, given the surprising requirement of heterogeneity to encode multiple intervals that is demonstrated. Moreover, we argue that the advantage of a heterogeneous distribution of time constants when minimizing the relative CRLB is minimal. Indeed, having an average relative CRLB of $106\%$ (largest heterogeneity used when $k=1$) instead of $128\%$ (homogeneous case when $k=1$) is not a practical advantage for such short time intervals (see Figure \ref{fig:continuous distribution}F).

That the best cell for decoding a single time interval be memory-less is intuitively satisfying. Indeed, this leaves all of the cell's resources to encode the latest time interval rather than being affected by the previous ones. Although a large proportion of the measured PG neurons had no memory, a significant number ($\sim 67 \%$) of them had $\beta > 0$~\cite{wallach_time-stamp_2018}. This suggests that a compromise is made between encoding the latest time interval and encoding the previous ones.

\subsection*{Benefit of negative current bias}

Lack of spontaneous (baseline) activity may seem to be a stringent requirement, yet most cells were silent between encounters ($c < 5$~spikes for $80\%$ of cells)~\cite{wallach_time-stamp_2018}. In fact, some of them had a negative value for this parameter $c$ (which acts like a bias), meaning they were inactive unless a strong enough stimulus could activate them, such as at a long interval. This may be a useful mechanism to encode a wider range of intervals in an efficient manner through adaptation. In fact, when $c \geq 0$, the only way a cell can encode larger intervals is by making $\tau$ larger, which moves the dynamic range. A silent cell ($c < 0$) can have a similar effect by moving this dynamic range while keeping a reasonable time constant. Even if $\tau$ is small relative to $T$, the cell can still encode the interval, since any activity implies the interval exceeds a minimal value set by $c$. A deeper analysis of this bias effect is beyond the scope of this paper.

\subsection*{Importance of heterogeneity for interval sequences}

Our evolving intuition was that more than one value of the recovery time constant is needed to optimally decode a wide range of time intervals, based on the finding that a specific value of $\tau$ minimizes the CRLB for a specific $T$ or distribution of $T$. However, averaging the CRLB over a simple prior distribution for $T$ does not yield additional optimal values of $\tau$. Instead, it shifts the minimum in a different direction. Although not proven here, our numerical explorations suggest that the CRLB is a convex function of $\left\{\tau^1, \tau^2, ...\right\}$, implying a global minimum. Since there is no difference in the expression of the CRLB from one population of neurons to the other for a given $T$, this global minimum must necessarily be at the point where $\tau^1=\tau^2=...$ Moreover, averaging over different values of $T$ simply has the effect of changing the position of this global minimum, because the sum of convex functions remains convex. This convexity argument makes it easier to understand how an homogeneous network is optimal when minimizing the average CRLB over a range of time intervals, no matter the prior distribution $P_T(T)$ (which can be seen as a positive weighted sum), since that operation preserves convexity and, therefore, the unique minimum. We also suspect that the error metric is not convex when looking at the relative CRLB for shorter time intervals, making their nonlinear combination when averaging over different time intervals generate additional minima. 

Heterogeneity in neural networks has been showed to encode a wide range of time scales~\cite{stern_reservoir_2023,marsat_cellular_2012}. It is also widely accepted as a good strategy to increase performance in different tasks, especially time-related ones~\cite{salaj_spike_2021}. The results presented for the single interval case suggest otherwise. The fact that recorded cells show a wide range of time constants and memory parameters~\cite{wallach_time-stamp_2018} then implies that the PG population is meant to encode sequences of more than one time interval. Moreover, the required heterogeneity proven here "in principle" is an important first step that needs to be followed by calculations of what that heterogeneity should be, including any correlated variability between parameters, to best encode interval sequences. This is not done in the present work, as an analysis as detailed as what is shown for a single time interval is greatly complexified by the augmented dimensionality of the problem upon considering multiple intervals.  

Since the estimation performance is similar when using different sensible distributions of time constants, it should be no surprise that past intervals can be encoded in adaptive responses without hindering the latest interval estimate too much. This is further confirmed by the fact that averaging over a wide range of time interval values renders the optimal $\tau$ less important. Indeed, any time constant that is in the proximity (usually larger) of this optimal value will perform similarly (see the flattening out of the curves past the optimal $\tau$ in Figure~\ref{fig:RMSEaverages}C). 

This can be expanded on when looking at the time resolution resulting from the combined activity of neurons in PG. If the time constant is large compared to the time interval between encounters, the activity of a single cell should be almost null. For example, suppose $\tau$ is large compared to $T$, such that PG neurons fire zero or one spike during an encounter due to almost no recovery. Not much can be said then from the activity of a single cell. However, the dynamic range (and therefore the time resolution) can be enhanced by downstream computation of the average activity of PG neurons, since the fraction of activated cells carries information about $T$. This explains why large non-optimal values of $\tau$ could reasonably encode smaller time intervals given a large population of neurons.

The analysis of adaptation-based single time interval encoding shown here provides a foundation for exploring the detailed multi-interval encoding. Adaptation processes with memory carry more information compared to those that reset upon spiking due to noise-shaping, which raises the possibility that the memory cells have a built-in noise reduction mechanism~\cite{nesse_biophysical_2010}. It is also of interest to extend our model to include synaptic plasticity, as that can enable single cell sequence anticipation and responses to unexpected inputs~\cite{saponati_sequence_2023}.

\section{Methods}\label{methods}

All numerical simulations are done in the \textit{Julia} programming language.

\subsection*{Computing the Fisher information and Cramér-Rao lower bound}

We first derive the expression for the FI with the general formulation. Recall that, for an estimator with log-likelihood (LL) $\ell(\left\{\theta_i\right\})$ and parameters $\left\{\theta_i\right\}$, the FI is given by
\begin{equation}
    \left[\mathcal{I}_{\theta}\right]_{ij} = -E\left[\frac{\partial^2\ell}{\partial\theta_i\partial\theta_j}\right],
\end{equation}
where $E[\cdot]$ is the expectation value over the observable data (here, the response of adaptive neurons). In the case of time interval estimation through MLE, the FI for one cell is given by
\begin{align}
    \left[\mathcal{I}_{T}\right]_{ij} &= -E_{R_n}\left[\frac{\partial}{\partial{T_i}}\left(\left(\frac{R_n}{\lambda_n} - 1\right)\frac{\partial \lambda_n}{\partial T_j}\right)\right] \nonumber,\\
    &= -E_{R_n}\left[ \left(\frac{R_n}{\lambda_n} - 1\right)\frac{\partial^2\lambda_n}{\partial T_i \partial T_j} - \frac{R_n}{\lambda_n^2}\frac{\partial\lambda_n}{\partial T_i}\frac{\partial\lambda_n}{\partial T_j}\right]\nonumber,\\
    &= \frac{1}{\lambda_n}\frac{\partial\lambda_n}{\partial T_i}\frac{\partial\lambda_n}{\partial T_j},
    \label{eq:FIderivation}
\end{align}
since $E_{R_n}[R_n] = \lambda_n$, by definition. For a network of multiple adaptive neurons, the FI simply becomes the sum of the FI of individual cells. This is due to the fact that the LL of the network is the sum of the LL of individual neurons and that the derivative and expectation value operators are linear. It is equivalently the consequence of assuming that the adaptive neurons in PG are independent. From this, we find the FI for the complete network to be given by Equation~\eqref{eq:fisherinfo}. The CRLB matrix is then given by the inverse of the FI matrix (see Equation~\eqref{eq:CRLB}). In the case of a single time interval, we obtain a scalar value for both. It's worth noting that, although going from a single to multiple cells in terms of FI can be done by simply adding to the FI matrix, the same is not true for the CRLB. The sum needs to be done prior to inverting the FI matrix.

\subsection*{Maximizing the Fisher information for a specific time interval}

For a single time interval, we can expand Equation~\eqref{eq:FIderivation} to (indices are dropped to lighten notation)
\begin{equation}
    \mathcal{I} = \frac{a^2(1-\beta x_0)^2e^{-2T/\tau}}{\tau^2\left[a(1 - e^{-T/\tau}(1 - \beta x_0)) + c\right]_+}.
    \label{eq:expanded_fisher_info}
\end{equation}
Assuming $c \geq 0$ (which is the case for $80\%$ of neurons in PG~\cite{wallach_time-stamp_2018}), the linear rectification can be dropped. To look at the behavior of the FI with respect to the parameters, we can look at its partial derivatives. First, looking at the derivative with respect to the baseline activity $c$, we obtain
\begin{equation}
    \frac{\partial \mathcal{I}}{\partial c} = -\frac{a^2(1-\beta x_0)^2 e^{-2T/\tau}}{\tau^2\left[a(1-e^{-T/\tau}) + c\right]^2}.
    \label{eq:dFIdc}
\end{equation}

For any admissible values of the parameters, Equation~\eqref{eq:dFIdc} is always negative. Therefore, the FI decreases as $c$ increases and can be maximized by making $c$ as small as possible (see Figure~\ref{fig:optimization of estimator}C). This means that the FI in this model deteriorates as the spontaneous activity increases. Since we assume $c \geq 0$, the optimal value for $c$ for any time interval value is $c = 0$.

We next look at the derivative with respect to the memory parameter $\beta$. When setting $c=0$, we have
\begin{align}
    \frac{\partial \mathcal{I}}{\partial \beta} = & -\frac{a(1 - \beta x_0)x_0e^{-2T/\tau}}{\tau^2\left[1 - e^{-T/\tau}(1 - \beta x_0)\right]} \times \nonumber \\ & \left[2 + \frac{(1 - \beta x_0)e^{-T/\tau}}{1 - e^{-T/\tau}(1 - \beta x_0)}\right].
    \label{eq:dFIdb}
\end{align}

Again, for any valid values of the parameters, Equation~\eqref{eq:dFIdb} is negative. The optimal memory parameter is therefore $\beta = 0$ when estimating a single time interval. After simplifying the expression with $c = \beta = 0$, it is straightforward to show that the derivative with respect to the gain parameter $a$ is always positive. Indeed, we have
\begin{equation}
    \frac{\partial \mathcal{I}}{\partial a} = \frac{e^{-2T/\tau}}{\tau^2(1 - e^{-T/\tau})}.
    \label{eq:dFIda}
\end{equation}

The FI is therefore unbounded with respect to $a$, i.e. a value of the gain as large as possible is optimal.

Finally, computing the derivative with respect to the recovery time constant $\tau$ yields a subtler result. In that case, we obtain the expression

\begin{equation}
    \frac{\partial \mathcal{I}}{\partial \tau} = -\frac{a}{\tau^3 \left(e^{2T/\tau} - e^{T/\tau}\right) } \left[ 2 - \frac{T\left(2 e^{2T/\tau} - e^{T/\tau}\right)}{\tau\left(e^{2T/\tau} - e^{T/\tau}\right)} \right].
    \label{eq:dFIdtau}
\end{equation}

Contrary to the previous parameters, the derivative of the FI with respect to $\tau$ is not always positive or negative. There is a maximum which can be found by setting the derivative to $0$. With a change of variable $x = T/\tau$ and using Newton's root finding algorithm, the optimum is found at the point $\tau \approx 1.55T$.

\subsection*{Computing the root mean square error of the estimator}

The need to be careful when using the FI and CRLB is well-known. This is because the FI is valid only for unbiased estimators. In the limit of an infinite number of responses, the estimator is expected to have exactly the same error as the CRLB. However, the minimum number of responses needed to effectively assume there is no bias in the estimator is not a trivial task~\cite{yarrow_fisher_2012}. We therefore need to compute the RMSE of the estimator with a Monte Carlo simulation and compare it to the CRLB. To do so, we generate the response of $N$ adaptive neurons resulting from encounters separated by an interval $T$. From these $N$ responses, we maximize the LL (see Equation~\eqref{eq:loglikelihood}) with respect to $T$ to find the estimate $T^{MLE}$. This process is repeated for a given number of samples $s$ such that the RMSE is given by

\begin{equation}
    RMSE(T) = \sqrt{\sum_{k=1}^{s} (T - T_k^{MLE})^2.}
    \label{eq:RMSE}
\end{equation}

Convergence is assumed when the difference in the average and variance of $T^{MLE}$ after $100$ new estimates is smaller than $10^{-6}$. This allows us to compare the actual error trend when estimating a single time interval (RMSE) with what we use to optimize the adaptation parameters (CRLB). For large enough networks, both quantities are essentially the same (see Results).

\subsection*{Averaging the Cramér-Rao lower bound over multiple time interval values}

When looking at a single value of time interval $T$, optimizing the FI is equivalent to optimizing the CRLB. However, the optimum becomes different when multiple values of time intervals are considered. To show that this is the case, we can simply look at the uniformly weighted sums for 2 time interval values $T_1^*$ and $T_2^*$ with a single type of cell for both the FI and the CRLB (see Equations \eqref{eq:FI comparison} and \eqref{eq:CRLB comparison}, respectively). The nonlinear sum of both quantities in Equation~\eqref{eq:CRLB comparison} causes the optimal $\tau$ to be at a different location than for Equation~\eqref{eq:FI comparison}:

\begin{align}
    \mathcal{I}_{\text{total}} &= \frac12\left(\mathcal{I}(T_1^*, \tau) + \mathcal{I}(T_2^*, \tau)\right), \label{eq:FI comparison}\\
    CRLB_{\text{total}} &= \frac12\left(\frac{1}{\sqrt{\mathcal{I}(T_1^*, \tau)}} + \frac{1}{\sqrt{\mathcal{I}(T_2^*, \tau)}}\right) \label{eq:CRLB comparison}.
\end{align}

Therefore, one has to choose for which of these two quantities the actual value of $\tau$ is optimal for. We argue that the choice of the CRLB is a natural one, since it lets us compare the actual error computed through Monte Carlo simulations to what is found from the FI. Moreover, for sequences of multiple time intervals, the CRLB of the total time travelled (sum of all time intervals in the sequence) is the sum of all elements in the inverted FI matrix (non-diagonal terms included), which makes it a convenient metric with which to optimize the cell parameters. Therefore, optimizing the CRLB makes the most sense. When adding new populations of cells and values of time intervals, the CRLB can easily be adapted to the general form

\begin{equation}
    CRLB_{\text{total}} = \sum_{k=1}^{n} \frac{q_k}{\sqrt{\sum_{j=1}^{N} p^j \mathcal{I}(T_k^*, \tau^j)}},
    \label{eq:general discrete CRLB}
\end{equation}

\noindent where $\sum_{k=1}^{n} q_k = \sum_{j=1}^{N} p^j = 1$. The weights $q_k$ allow the assignment of importance to specific values of time intervals, while the weights $p^j$ represent the proportion of cells with time recovery parameter $\tau^j$ in the network of PG cells. The continuous limit of this equation was used in the optimization of the parameters (see Equation~\eqref{eq:avgCRLB}). The integral was computed numerically with an adaptive Gauss-Kronrod integration technique from the \textit{QuadGK.jl} library and was then optimized with the L-BFGS method from the \textit{Optim.jl} library while keeping the value of the average $\tau$ between $0.1$~s and $80$~s (box-constrained).

\subsection*{Computing the determinant for a sequence of 2 time intervals}

The FI matrix needs to be invertible to have a finite lower bound on the error made on the estimates of time intervals. For two populations of sizes $N^1$ and $N^2$, the determinant of the FI matrix is given by

\begin{align}
    \text{det}\mathcal{I} = &\left[ \frac{N^1}{\lambda_2^1}\left(\frac{\partial \lambda_2^1}{\partial T_1}\right)^2 + \frac{N^2}{\lambda_2^2}\left(\frac{\partial \lambda_2^2}{\partial T_1}\right)^2 \right] \times \nonumber \\ &
                             \left[ \frac{N^1}{\lambda_2^1}\left(\frac{\partial \lambda_2^1}{\partial T_2}\right)^2 + \frac{N^2}{\lambda_2^2}\left(\frac{\partial \lambda_2^2}{\partial T_2}\right)^2 \right] \nonumber \\
                             & - \left[ \frac{N^1}{\lambda_2^1}\left(\frac{\partial \lambda_2^1}{\partial T_1}\right)\left(\frac{\partial \lambda_2^1}{\partial T_2}\right) \right. \nonumber \\ & \left. \qquad + 
                                       \frac{N^2}{\lambda_2^2}\left(\frac{\partial \lambda_2^2}{\partial T_1}\right)\left(\frac{\partial \lambda_2^2}{\partial T_2}\right) \right]^2.
                          \label{eq:full_det_step1}
\end{align}

Expanding the multiplication of the terms in brackets yields
\begin{align}
    \text{det}\mathcal{I} &= \frac{(N^1)^2}{(\lambda_2^1)^2}\left(\frac{\partial \lambda_2^1}{\partial T_1}\right)^2\left(\frac{\partial \lambda_2^1}{\partial T_2}\right)^2 
                             + \frac{N^1N^2}{\lambda_2^1\lambda_2^2}\left(\frac{\partial \lambda_2^2}{\partial T_1}\right)^2\left(\frac{\partial \lambda_2^1}{\partial T_2}\right)^2  \nonumber\\
                             &\quad + \frac{(N^2)^2}{\lambda_2^2}\left(\frac{\partial \lambda_2^2}{\partial T_1}\right)^2\left(\frac{\partial \lambda_2^2}{\partial T_2}\right)^2 
                             + \frac{N^1N^2}{\lambda_2^1\lambda_2^2}\left(\frac{\partial \lambda_2^1}{\partial T_1}\right)^2\left(\frac{\partial \lambda_2^2}{\partial T_2}\right)^2 \nonumber\\
                             &\quad - \frac{(N^1)^2}{(\lambda_2^1)^2}\left(\frac{\partial \lambda_2^1}{\partial T_1}\right)^2\left(\frac{\partial \lambda_2^1}{\partial T_2}\right)^2 
                             - \frac{(N^2)^2}{(\lambda_2^2)^2}\left(\frac{\partial \lambda_2^2}{\partial T_1}\right)^2\left(\frac{\partial \lambda_2^2}{\partial T_2}\right)^2 \nonumber\\
                             & \quad - 2\frac{N^1N^2}{\lambda_2^1\lambda_2^2}\left(\frac{\partial \lambda_2^1}{\partial T_1}\right)\left(\frac{\partial \lambda_2^1}{\partial T_2}\right)\left(\frac{\partial \lambda_2^2}{\partial T_1}\right)\left(\frac{\partial \lambda_2^2}{\partial T_2}\right).
                             \label{eq:full_det_step_2}
\end{align}

Many terms can be cancelled and, putting $N^1N^2/\lambda_2^1\lambda_2^2$ in front of the expression and splitting the last term of Equation~\eqref{eq:full_det_step_2} in two, we obtain
\begin{align}
\text{det}\mathcal{I} =  \frac{N^1N^2}{\lambda_2^1\lambda_2^2}&\left[\left(\frac{\partial \lambda_2^2}{\partial T_1}\right)^2\left(\frac{\partial \lambda_2^1}{\partial T_2}\right)^2 \right. \nonumber \\ & \left. - \left(\frac{\partial \lambda_2^1}{\partial T_1}\right)\left(\frac{\partial \lambda_2^1}{\partial T_2}\right)\left(\frac{\partial \lambda_2^2}{\partial T_1}\right)\left(\frac{\partial \lambda_2^2}{\partial T_2}\right) \right. \nonumber\\
                          &\left. + \left(\frac{\partial \lambda_2^1}{\partial T_1}\right)^2\left(\frac{\partial \lambda_2^2}{\partial T_2}\right)^2 \nonumber \right. \\ & \left.- \left(\frac{\partial \lambda_2^1}{\partial T_1}\right)\left(\frac{\partial \lambda_2^1}{\partial T_2}\right)\left(\frac{\partial \lambda_2^2}{\partial T_1}\right)\left(\frac{\partial \lambda_2^2}{\partial T_2}\right) \right].
                          \label{eq:full_det_step3}
\end{align}

Then, we factor one of the squares to get
\begin{align}
\text{det}\mathcal{I} = \frac{N^1N^2}{\lambda_2^1\lambda_2^2}&\left[\left(\frac{\partial \lambda_2^2}{\partial T_1}\right)\left(\frac{\partial \lambda_2^1}{\partial T_2}\right)\left\{\left(\frac{\partial \lambda_2^2}{\partial T_1}\right)\left(\frac{\partial \lambda_2^1}{\partial T_2}\right) \right. \right. \nonumber \\ & \qquad \qquad \qquad \qquad \left. \left. - \left(\frac{\partial \lambda_2^1}{\partial T_1}\right)\left(\frac{\partial \lambda_2^2}{\partial T_2}\right)\right\} \right. \nonumber\\
                          &\left. + \left(\frac{\partial \lambda_2^1}{\partial T_1}\right)\left(\frac{\partial \lambda_2^2}{\partial T_2}\right)\left\{\left(\frac{\partial \lambda_2^1}{\partial T_1}\right)\left(\frac{\partial \lambda_2^2}{\partial T_2}\right) \right. \right. \nonumber \\ & \qquad \qquad \qquad \qquad \left. \left. - \left(\frac{\partial \lambda_2^2}{\partial T_1}\right)\left(\frac{\partial \lambda_2^1}{\partial T_2}\right)\right\}\right].
                          \label{eq:full_det_step4}
\end{align}

Finally, we note that the terms in the curly brackets can be factored out to retrieve the result shown in Equation~\eqref{eq:FIdet}. This shows the necessity of at least two types of cell - i.e. of parametric heterogeneity -  to encode two successive time intervals.

\bmhead{Supplementary information}

A PDF document with supplementary derivations and figure is attached to the submission of this paper.

\bmhead{Acknowledgements}

This work was supported by NSERC grant RGPIN/06204-2014 to AL and by FRQ grant B2X/328560 to RLM. 

\bmhead{Author contributions}

RLM did the theoretical work. RLM and AW performed analytical calculations. RLM performed numerical simulations. RLM, LM, AW and AL contributed to the interpretation of the results. RLM and AL wrote the manuscript. LM and AW edited the manuscript.

\bmhead{Competing interests}

The authors declare no competing interests.





\bibliography{sn-bibliography}

\end{document}